\documentclass[useAMS,usenatbib,usegraphicx,psfig]{mn2e}

\title[]{The 2018 eruption and long term evolution of the new high-mass Herbig Ae/Be object Gaia-18azl = VES 263}
\author[]{U. Munari$^{1}$, V. Joshi$^{2}$, D.P.K. Banerjee$^{2}$, K. {\v C}otar$^{3}$, Shugarov S.Y.$^{4}$,
\newauthor R. Jurdana-{\v S}epi{\'c}$^{5}$, R. Belligoli$^{6}$, A. Bergamini$^{6}$, M. Graziani$^{6}$,
\newauthor G.L. Righetti$^{6}$, A. Vagnozzi$^{6}$, P. Valisa$^{6}$\\
\\
$^{1}$INAF Astronomical Observatory of Padova, 36012 Asiago (VI), Italy\\
$^{2}$Physical Research Laboratory, Navrangpura, Ahmedabad, Gujarat 380009, India\\
$^{3}$Faculty of Mathematics and Physics, University of Ljubljana, Jadranska 19, 1000 Ljubljana, Slovenia\\
$^{4}$Sternberg Astronomical Institute, M.V. Lomonosov Moscow State University, Moscow 119991, Russia\\
$^{5}$University of Rijeka, Physics Department, Radmile Matej{\v c}i{\'c}, 51000, Rijeka, Croatia\\
$^{6}$ANS Collaboration, Astronomical Observatory, 36012 Asiago (VI), Italy\\}

\begin{document}

\maketitle

\label{firstpage}

\begin{abstract}
We have been  monitoring, at high cadence,  the photometric and
spectroscopic evolution of VES 263 following the discovery in 2018 of a
brightening labeled as event Gaia-18azl.  VES 263 is so far a neglected
emission-line object discovered in the 1960s on objective prism plates,
tentatively classified as a semi-regular AGB cool giant by automated
analysis of ASASSN lightcurves.  We have discovered that VES 263 is a
bonafide massive pre-Main Sequence object ($\sim$12 M$_\odot$), of the
Herbig AeBe type.  It is located at 1.68$\pm0.07$ kpc distance, within the Cyg OB2
star-forming region, and it is highly reddened ($E_{B-V}$=1.80$\pm$0.05) by
interstellar extinction.  In quiescence, the spectral energy
distribution is dominated by the $\sim$20,000~K photospheric emission from
the central B1II star, and at $\lambda$$\geq$6$\mu$m by emission from
circumstellar warm dust ($T$$\leq$400$^\circ$K). The 2018-19 eruption was
caused by a marked brightening of the accretion disk around the B1II star as
traced by the evolution with time of the integrated flux and the
double-peaked profile of emission lines.  At the peak of the eruption, the
disk has a bulk temperature of $\sim$7500~K and a luminosity
$L\geq$860~L$_\odot$, corresponding to a mass accretion rate
$\geq$ 1.1$\times10^{-5}$~M$_\odot$ yr$^{-1}$.  Spectroscopic signature of
possible bipolar jets (at $-$700 and $+$700 km~s$^{-1}$) of variable
intensity are found.  We have reconstructed from Harvard, Moscow and
Sonneberg photographic plates the photometric history of VES 263 from 1896
to 1995.
\end{abstract}

\begin{keywords}
stars: pre-main-sequence; stars: variables: T Tauri, Herbig Ae/Be;
open clusters and associations: individual: Cyg OB2
\end{keywords}

\section{Introduction}

On April 18, 2018 the Gaia team issued an alert for a 0.5 mag increase
observed by Gaia on a field star located at (J2000) $\alpha$=20$^h$ 31$^m$
48$^s$.85, $\delta$=+40$^\circ$ 38' 00''.1, rising from a previous average
of $G$=12.16 to $G$=11.66 mag.  The brightening star was logged as Gaia-18azl,
and the event was also filed as AT-2018awf by the IAU Transient Name Server. 
On this server, a couple of days past discovery, R.  Fidrich noted the
presence at this position of an anonymous star recognized as a variable by
the ASASSN sky patrol (and logged as ASASSN-V J203148.85+403800.1). 
Fridrich commented as the ASASSN automated classification of the star as a
semi-regular variable of P=197 days seemed unlikely given the ASASSN
lightcurve, and suggested instead it being a young stellar object in
outburst.

A positional search through the literature revealed Gaia-18azl being
identical with VES 263, a poorly known emission-line star catalogued by the
Vatican Emission Star survey (Wisniewski and Coyne 1976), and by Stephenson
and Sanduleak (1977) as SS 447.  The latter noted strong H$\alpha$ in
emission, but said nothing about the underlying continuum.  The latter was
classified as that of a B star by Downes and Keyes (1988) from spectroscopic
observations but with no further details or a spectrum being shown.  The
object was also observed as having the H$\alpha$ line in emission in the
course of the Hamburg Observatory objective-prism sky survey (carried out in
1964-70), entering the Kohoutek and Wehmeyer (1997, 1999) catalog as HBH$\alpha$
4203-31.  Comer{\'o}n and Pasquali (2012, hereafter CP12), in the course of
a survey of the Cygnus OB2 association and its surroundings in search for
previously unrecognized massive stars, obtained a low resolution spectrum of
the same object, logged as J20314885+4038001, without commenting on its
positional coincidence with entries in catalogs of emission-line stars. 
They classified the spectrum as a highly reddened B1II star and in their
plot, covering the range 3900-4900\AA~ (i.e.  H$\epsilon$-H$\beta$), no
emission line is readily evident, including H$\beta$ that looks in normal
absorption without an emission core.

VES 263 lies positionally close to the center of the massive Cygnus OB2
association, at the outskirts of the $\gamma$ Cyg bright nebula and the dark
cloud LDN 889 (Lynds 1962) which appears superimposed to it.  Inspection of
IRAS mid-IR maps shows VES 263 being at the very heart of a highly
structured, knotty dust-complex extending for $\sim$2$^\circ$ in radius. 
This is an area of intensive star formation, with the two highest mass
Herbig Ae/Be stars lying close to VES 263 viz.  MWC 1021 at 72 arcmin and
V1478 Cyg at 14 arcmin.  Herbig Ae/Be stars (HAeBes for short) are pre-main
sequence stars of intermediate mass, spanning the range between the
lower-mass T Tauri stars and the embedded massive young stellar objects
(MYSOs).  The most recent catalog of HAeBes by Vioque et al.  (2018), lists
252 entries, with only 79 of them corresponding to B- or O-spectral types
and the rest being cooler/lower-mass stars.  The discovery of any new
high-mass HAeBe object is therefore relevant given the limited number of
such objects known in the whole Galaxy.

    \begin{figure}
    \includegraphics[width=80mm]{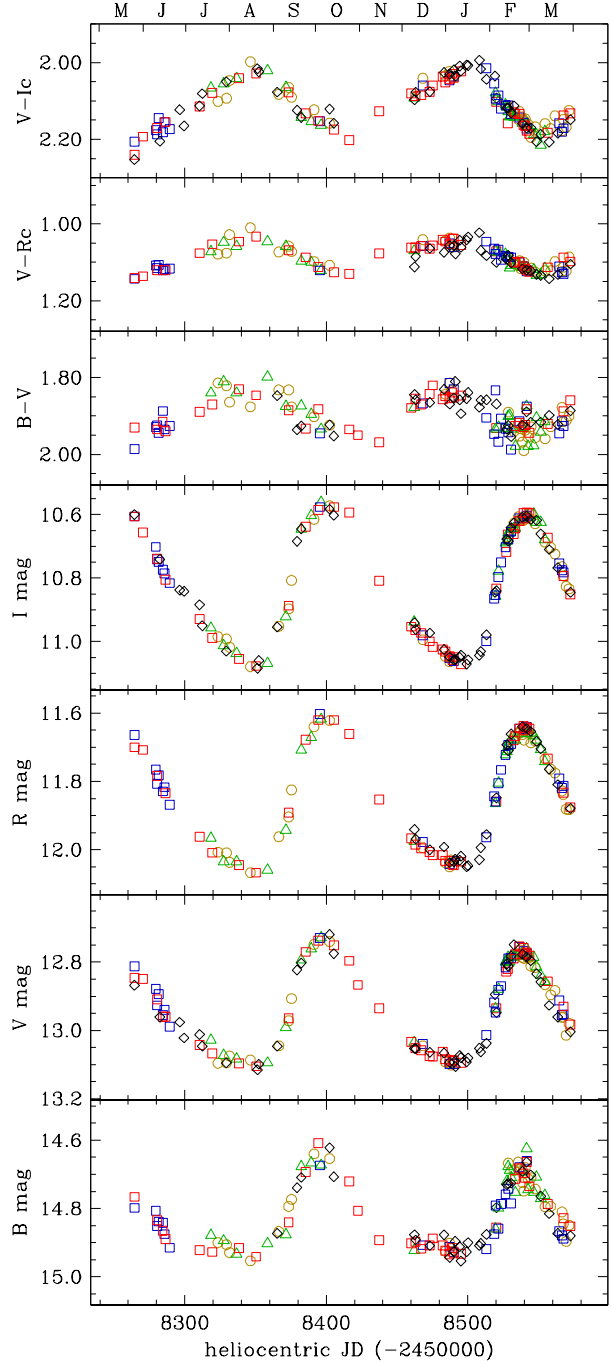}
    \caption{The $B$$V$$R_{\rm C}$$I_{\rm C}$ light- and color-evolution of
    the 2018-19 outburst of Gaia-18azl = VES 263 from our observations
    listed in Table~1. Data provided by different ANS Collaboration
    telescopes are marked with different symbols. Letters at the top mark
    the month from May 2018 to March 2019.}
    \end{figure}

The B1II spectral type, the strong H$\alpha$ feature in emission, the
partnership with a site of ongoing stellar formation and the location at the
boundaries of a dark interstellar cloud perfectly match VES 263 with the
classification criteria formulated by Herbig (1960) for HAeBe objects. 
Motivated by this, after the Gaia alert we initiated a high cadence
$B$$V$$R_{\rm C}$$I_{\rm C}$ monitoring, later augmented by low- and
high-resolution optical spectroscopy, near-IR observations and a search
through photographic archival repositories, the results of which are
presented and discussed here.  We will show how VES 263 is indeed a bona
fide new HAeBe star, heavily obscured by interstellar extinction and
with IR excess due to circumstellar dust, showing a complex century-long
photometric history and a marked variability rapidly increasing toward the
Infrared, in addition to a bright circumstellar disk where double-peaked
emission lines form and which is responsible for the recorded photometric
variability.

    \begin{figure}
    \includegraphics[width=85mm]{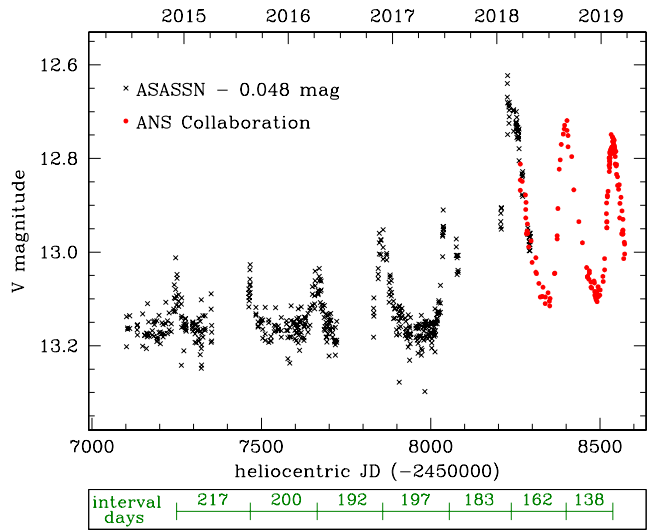}
    \caption{The recent photometric history of VES 263 constructed by combining
    archival $V$-filter photometry from the ASASSN project with current outburst
    $V$-band data from the ANS Collaboration. The latter are fully color
    transformed to the Landolt system, while ASASSN are not and require the
    indicated $-$0.048 mag shift to be brought into agreement with APASS and
    ANS photometry.}
    \end{figure}

    \begin{figure}
    \includegraphics[width=85mm]{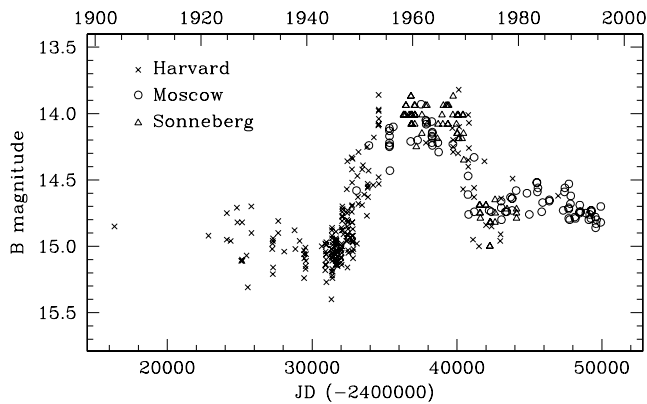}
    \caption{Historical blue-band lightcurve of VES 263 constructed  from
    Harvard, Moscow and Sonneberg photographic plates.}
    \end{figure}

\section{Observations}

\subsection{Optical photometry}

$B$$V$$R_{\rm C}$$I_{\rm C}$ photometry (in the Landolt system) of VES 263
was obtained with five ANS Collaboration telescopes (with identifiers 606,
703, 1301, 1507, and 2900), all located in Italy and each working
independently from all others (for simplicity, in the rest of the paper we
will adopt $R$ and $I$ for band nomenclature instead of the more precise
$R_{\rm C}$ and $I_{\rm C}$).  The operation of ANS Collaboration telescopes
is described in detail by Munari et al.  (2012) and Munari \& Moretti
(2012).  The observations at each site were transformed from the
instantaneous local photometric system to the standard Landolt system via a
common local photometric sequence established around VES 263 and covering a
wide range in color that brackets that of the program star.  An initial
version of the local photometric sequence was extracted from the APASS all
sky survey (e.g.  Henden and Munari 2014) and ported to Landolt equatorial
system via the transformations calibrated in Munari et al.  (2014).  The
sequence was then continuously improved as a by-product of the accumulating
observations of VES 263, and by the end of the campaign, all observations
were re-reduced on the central ANS Collaboration server against this
improved and final sequence.  In all we collected 163 independent
$B$$V$$R$$I$ runs, on 109 different nights, from 2018 May 25 to 2019 March
30.  The results are listed in Table~1 (available in full in electronic form
only).  The total error budget (quadratic sum of the Poissonian noise on the
variable and the formal error on transformation from the local instantaneous
system to the standard one via color equations) has a median value of 0.009
mag for the data in Table~1 and it is therefore omitted.  The resulting
color-and light-curves are plotted in Figure~1.

\subsection{Optical spectroscopy}

Low, medium and high resolution spectra of VES 263 were obtained with three
different telescopes located in Asiago and Varese (Italy).  In all
observations the spectrograph slit was aligned along the parallactic angle
for optimal flux calibration, achieved by comparison against a
spectrophotometric standard located close to VES 263 in the sky and observed
either immediately before or after the target.  Data reduction was carried
out in IRAF and involved the usual steps on bias and dark removal,
flat-fielding, variance-weighted spectrum tracing, sky subtraction,
wavelength calibration, heliocentric correction and flux calibration.  A log
of the spectroscopic observations at optical wavelengths is provided in
Table~2.

The Asiago 1.22m deployed  a B\&C spectrograph and ANDOR iDus DU440A
camera housing a back-illuminated E2V 42-10 CCD (2048$\times$512 array, 13.5
$\mu$m pixels).  The low dispersion observations of VES 263 were obtained
with a 300 ln/mm grating blazed at 5000 \AA, providing a dispersion of 2.31
\AA/pix over the range $\lambda\lambda$~3300$-$8000~\AA, while medium
dispersion observations used two 1200 ln/mm gratings, one blazed at 4000
\AA\ and the other at 6500 \AA\, both giving a dispersion of 0.60
\AA/pix.

On the Asiago 1.82m telescope,  to record the high resolution spectra of VES
263, we used the REOSC-Echelle spectrograph, which is equipped with an EEV
CCD47-10 CCD (1024$\times$1024 array, 13 $\mu$m pixels).  It covers the
interval $\lambda\lambda$~3700$-$7300~\AA\ in 30 orders without inter-order
gaps.  A slit width of 1.5 arcsec was used resulting in a resolving power of
20,000.

The Varese 0.84m telescope was used in conjunction with a mk.III Multi Mode
Spectrograph from Astrolight Instr., feeding light to a SBIG ST-10XME camera
(2184$\times$1472 array, 6.8 $\mu$m).  The slit width was allowed to be
variable between 2 and 3 arcsec in accordance with the seeing.  In the
Echelle configuration the Multi Mode Spectrograph covers the range
$\lambda\lambda$~4225$-$8910~\AA\ in 27 orders, at a resolving power of
16,000 and without significant inter-order wavelength gaps.  In the low
dispersion mode, a 600 ln/mm grating blazed at 5000 \AA\ was adopted
allowing coverage of the interval $\lambda\lambda$~4585$-$9490~\AA\ at 2.10
\AA/pix.

   \begin{table}
   \caption{Optical photometry of VES 263 (a small portion is shown here to
   provide guidance about its content: the full table is available in
   electronic form only). The last column lists the telescope identifier,
   the same as used in Figure~1.}
   \includegraphics[width=70mm]{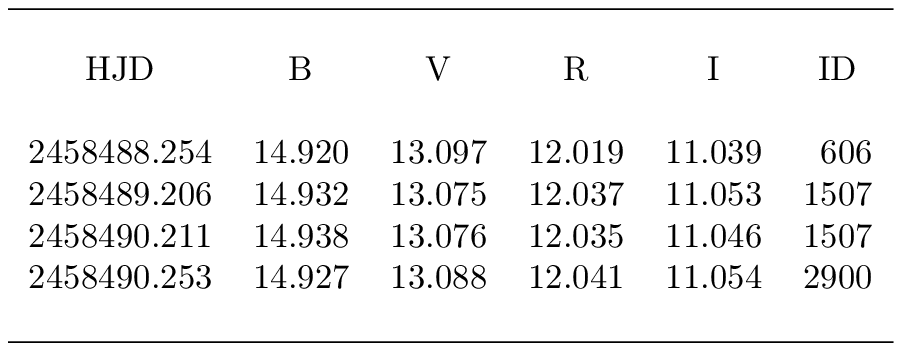}
   \end{table}

   \begin{table}
   \caption{Log of optical spectroscopy. The last column lists the integrated
   absolute flux of H$\alpha$ (in units of 10$^{-13}$ erg/cm$^2$~sec).
   The symbol $\ddag$ marks nights without observations of
   spectro-photometric standard stars.}
   \includegraphics[width=85mm]{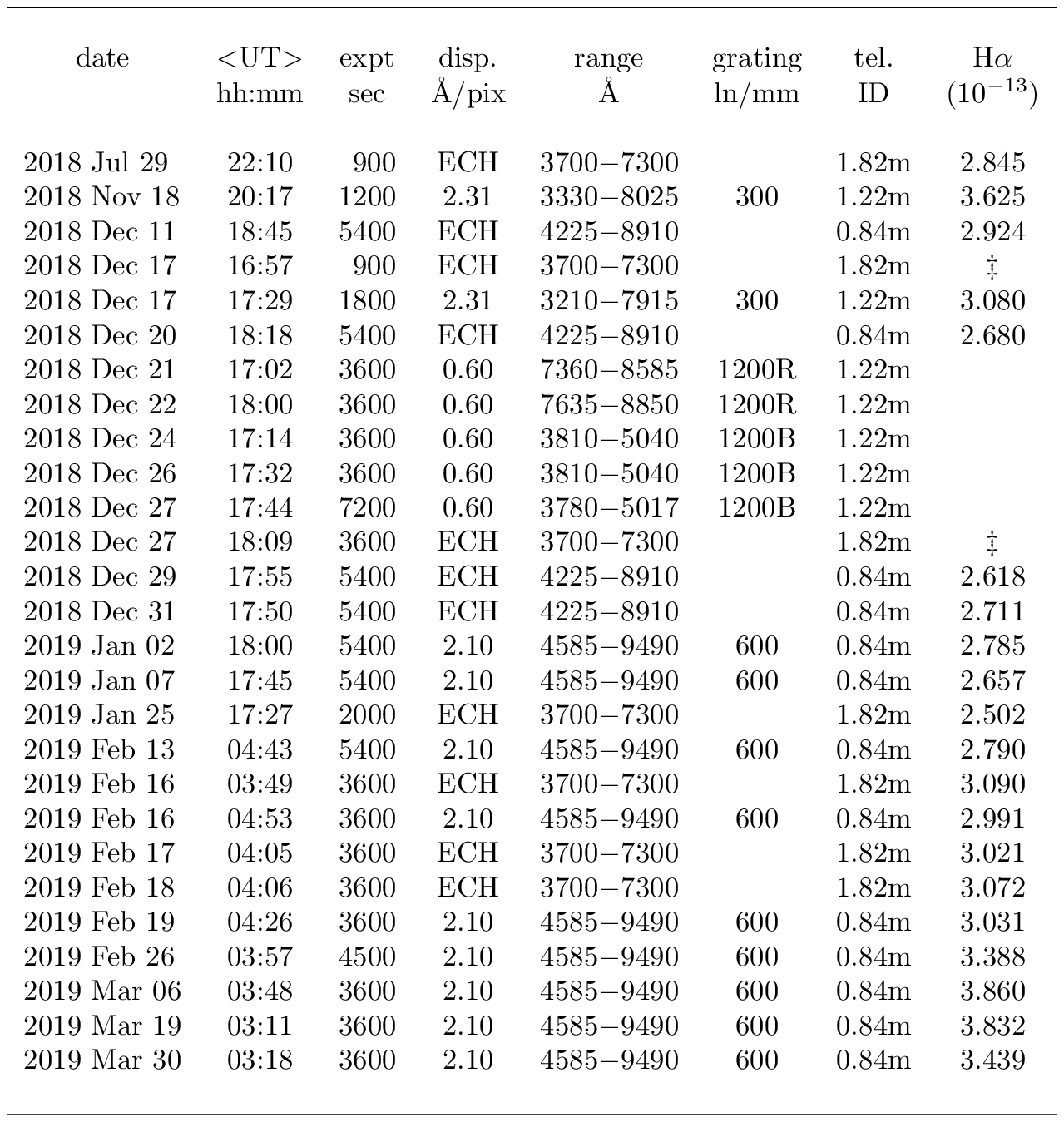}
   \end{table}

   \begin{table}
   \caption{Infrared photometry of VES 263.}
   \includegraphics[width=75mm]{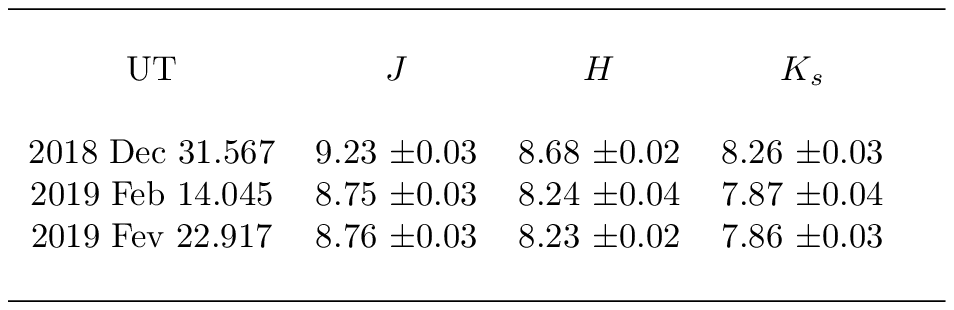}
   \end{table}

   \begin{table}
   \caption{$B$-band magnitude of VES 263 estimated on photographic plates
   taken with astrographs in Crimea (C) and Sonneberg (S).  The JD column
   lists the geocentric JD - 2400000 (a small portion is shown here to provide 
   guidance about its content: the full table is available in electronic form
   only).}
   \includegraphics[width=30mm]{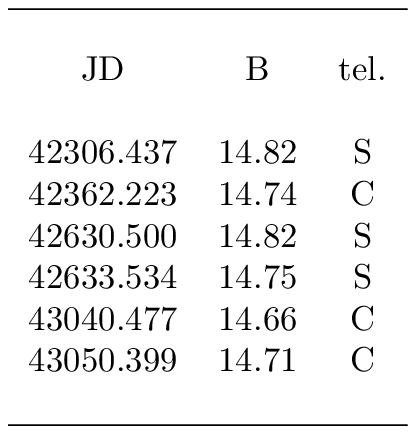}
   \end{table}

    \begin{figure*}
    \includegraphics[width=170mm]{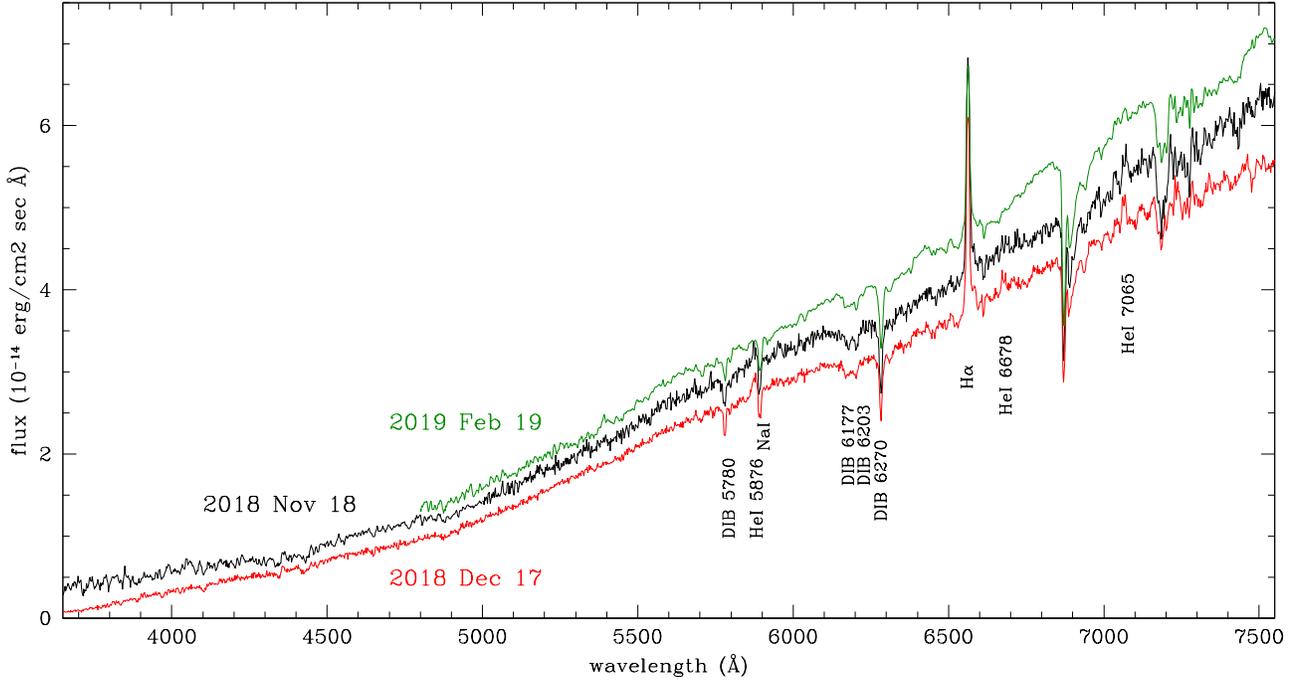}
    \caption{Sample low resolution spectra of VES 263 for three distinct
    states during the 2018-19 eruption: minimum (2018 Dec 17), peak (2019
    Feb 19) and intermediate brightness (2018 Nov 18).  The strongest
    interstellar absorption features and stellar emission lines are
    identified.}
    \end{figure*}

    \begin{figure*}
    \includegraphics[width=170mm]{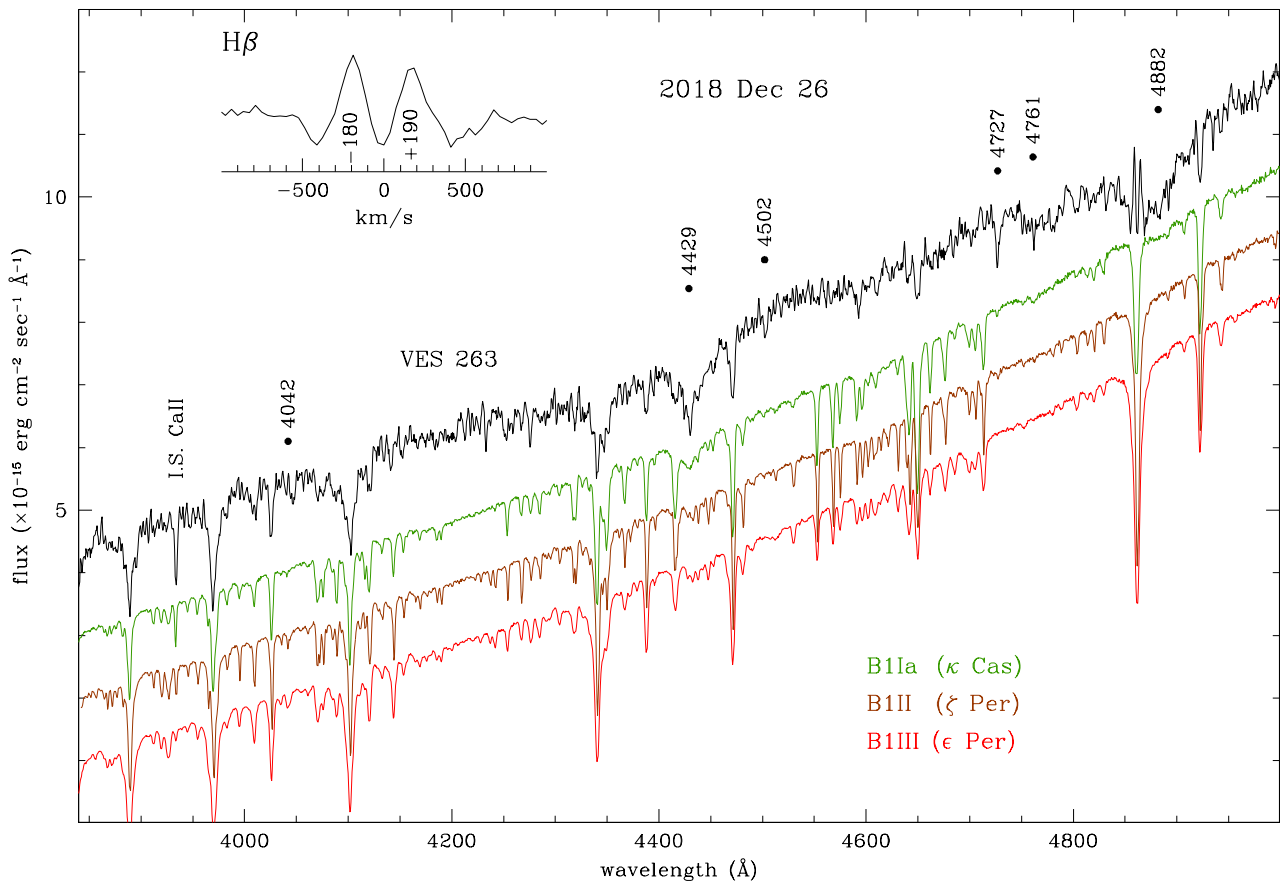}
     \caption{Medium resolution spectrum (0.60 \AA/pix) of VES 263 for 2018
     Dec 26 compared to those of MKK standard stars observed with the same
     instrumental set-up.  The standards have been scaled to the same flux
     and slope of VES 263 and have offsets applied  for plot clarity.  The inset
     shows the velocity profile of the  H$\beta$ double-peaked emission of VES
     263.  Dots mark the strongest diffuse insterstellar bands.}
    \end{figure*}

\subsection{Near-IR}

NIR spectroscopy and $JHK_s$ band photometry of VES 263 were obtained at a
few epochs with the 1.2m telescope of the Mount Abu Infrared Observatory
(Banerjee \& Ashok 2012) using the Near-Infrared Camera/Spectrograph
equipped with a 1024$\times$1024 HgCdTe Hawaii 1 array.

In photometry, the camera provides an unvignetted 8$\times$8 square arcmin
field-of-view.  Frames in each filter were obtained in five dithered
positions offset typically by 30 arcsec, with 5 frames being obtained in
each dithered position.  The corrected science frames were again median
combined to produce a sky frame which was subtracted from the individual
science frames.  Flat field correction was applied using a sky flat derived
from dark subtracted raw frames.  Finally, the frames were corrected for bad
pixels and cosmic ray hits.  The final corrected science frames were
co-added to produce an average frame on which aperture photometry was done
using routines in IRAF with the 2MASS field stars J20314618+4041107,
J20321252+4039474 and J20320263+4035518 used for photometric calibration.
The recorded JHKs photometry of VES 263 is given in Table~3.

A near-IR spectrum of VES 263 was obtained on 2019 December 31.603 UT at an
airmass $\sim$2.7, covering the wavelength range 0.90 to 1.78$\mu$m and at a
resolution R$\sim$1000.  The star was dithered to two positions along the
slit to allow for sky and dark subtraction.  These sky-subtracted images
were used to extract 1D spectra.  Wavelength calibration, accurate to a
1$\sigma$ value of 0.0002 $\mu$m, was done using a combination of OH airglow
lines and telluric lines that register with the stellar spectra.  To remove
telluric lines and to correct for instrumental response function, the target
spectra were ratioed with the spectral standard SAO 71278 (Sp.  type A0V,
Teff = 9750 K), from whose spectra the hydrogen Paschen and Brackett
absorption lines had been removed.  The blackbody curve corresponding to the
effective temperature of the spectral standard star was finally multiplied
with the ratioed spectra.  Data reduction and analysis was done using IRAF
tasks and Python routines developed by us.

\subsection{Objective prism plates}

Six deep 103a-F photographic plates, exposed through a 4$^{\circ}_{.}$5
objective prism (650 \AA/mm at H$\gamma$) on various nights of November
1973, were found in the plate archive of the Asiago 67/92cm Schmidt
telescope covering the sky around VES 263.  Eye inspection through a high
quality binocular Zeiss microscope revealed  H$\alpha$ to be in strong
emission in VES 263 at that time.  The underlying continuum appears very red
and the contrast of the H$\alpha$ emission is compatible with a photographic
rendition of the CCD spectrum shown in Figure~4.

\section{The 2018-19 eruption}

    \begin{figure*}
    \includegraphics[width=170mm]{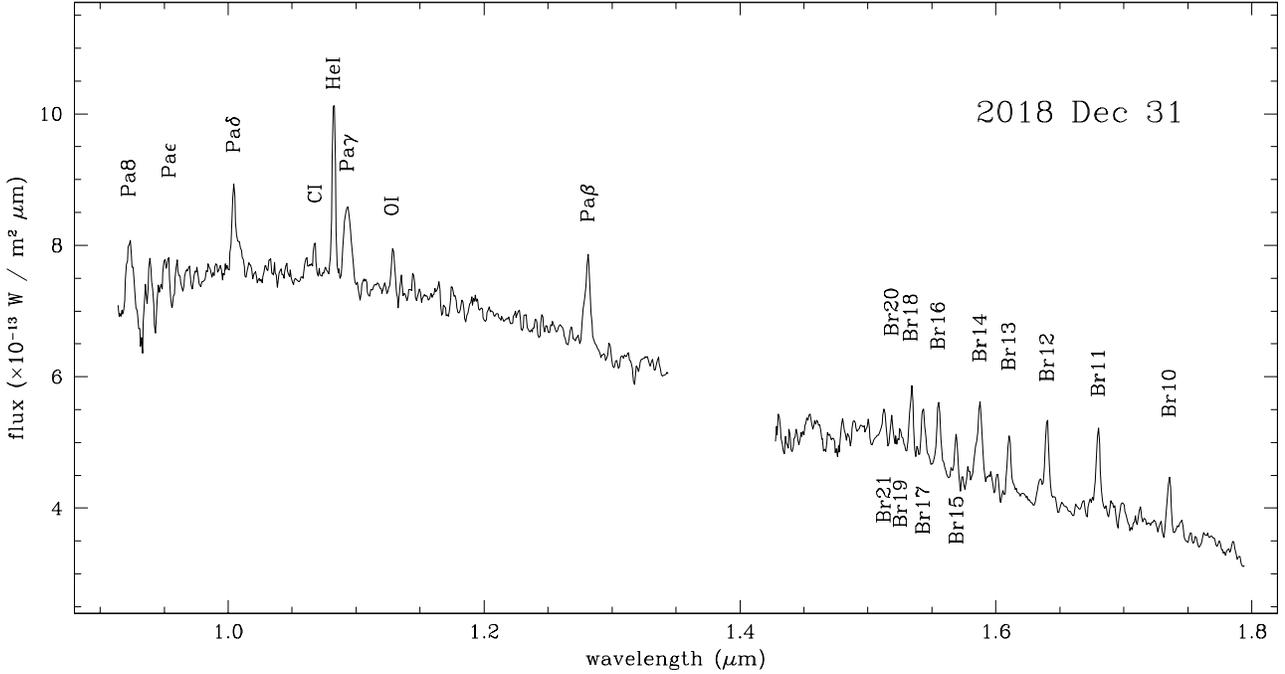}
    \caption{Near-IR spectrum of VES 263 for 2018 Dec 31 covering the $Y$,
    $J$, and $H$ band wavelength range. The strongest emission lines are
    identified.}
    \end{figure*}

\subsection{Photometric behaviour}

Our {\it BVRI} photometric monitoring of the 2018-19 eruption of VES 263,
triggered by the Gaia-18azl alert, is presented in Figure~1.  The behaviour is
very smooth (no noticeable short time-scale changes), with the amplitude of
variation growing with increasing wavelength, from $\Delta B$=0.31 mag to
$\Delta I$=0.50 mag.

During the current 2018-19 eruption, VES 263 is redder than in the
preceding quiescence, and the variability within the eruption is
characterized by redder colors when the star is brighter and bluer when
fainter, in a tight correlation.  This color behaviour precludes either a
variable dust obscuration or a change in photospheric temperature affecting
the B1II star as the causes of the observed variability.  An increase in
photospheric temperature would in fact make VES 263 brighter but also bluer
over the whole $\lambda$-range.  Similarly a reduction in the amount of dust
extinction, if any should occur, would result in a brighter VES 263, but
again in bluer colors.

To put the data for 2018-19 eruption into perspective, we have retrieved
patrol data from the ASASSN archive (Shappee et al.  2014, Kochanek et al. 
2017) and use them in Figure~2 to built a longer baseline 2014-2019 $V$-band
lightcurve for VES 263.  The ASASSN $V$-filter photometry is derived
differentially with respect to field stars, and no color-transformation to
the standard $V$-band is performed (M.  Pawlak private communication).  This
may be relatively inconsequential for stars of neutral colors, but it leads
to appreciable offsets for very blue or very red objects.  Comparing APASS
and ASASSN photometry for VES 263 in quiescence, it is obvious how the
ASASSN $V$-filter data needs the application of a $-$0.048 mag offset to be
brought into agreement with proper $V$-band magnitudes.  The same offset is
obtained when comparing ASASSN and our data for the current outburst. 
Therefore a $-$0.048 offset is applied to the ASASSN data before plotting
them in Figure~2.  The formal internal error of ASASSN data ranges
from 0.011 to 0.067, with a median value of 0.015 mag.

The 2014-2017 part of the lightcurve in Figure~2 shows VES 263 stable at
$V$=13.15, with small-amplitude and short-lived brightenings by $\Delta
V$~0.1/0.2 mag.  The following three brightenings, starting with the one
noticed by Gaia at the beginning of 2018, are larger in amplitude and
characterized by the distinctive feature that at minima the system remains
brighter than in quiescence.  As illustrated by the bar at the bottom of
Figure~2, the time-interval between successive brightenings steeply
decreases from 217 to 138 days.  This precludes them  being caused by
changing orbital aspect in a binary system or by rotation of a heavily
spotted photosphere.

The steep decrease in the time-interval between successive brightenings also
argues against them being caused by radial pulsations, in spite of the shape
of the lightcurve in Figure~1 reminiscent of Cepheid variables (and
similarly for RR Lyr and Mira types).  Pulsating variable are known to
change in amplitude and period, but much more gradually and by smaller
amounts (e.g.  Sterken \& Jaschek 1997).  However, common to all types of
radial pulsators is the fact that their colors get {\em bluer} on the rise
in brightness, while VES 263 instead turns {\rm redder}.  Furthermore, VES
263 lies far away from the radial pulsation instability strip on the HR
diagram (see sect.7 below), and the types of non-radial pulsating variables
observed around its position on the HR diagram (i.e.  53 Per, $\alpha$ Cyg,
$\beta$ Cep) are characterized by (much) shorter periods and smaller
amplitudes (as documented in the {\it General Catalog of Variable Stars} by
Samus' et al.  2017).

In this section we have been able to exclude various possible causes for the
variability and current eruption displayed by VES 263 viz. variable dust
obscuration, temperature changes, pulsations, stellar rotation and orbital
motion.  The agent responsible for the observed variability seems therefore
distinct from the B1II star, something appreciably cooler, and of a large
size in order to match its radiation output. In the following we'll argue
this is caused by a circumstellar disk.

\subsection{Long term evolution from historic photographic plates}

No variability of VES 263 was noted before the recent Gaia trigger and the
ASASSN patrol data, in spite of the convenient sky location and apparent
brightness of the object.  The obvious question thus concerns how unique is
the present eruption in the context of the recorded history of the object. 
To investigate its past photometric behaviour, we turned our attention to
the photographic plate stacks at Harvard, Moscow and Sonneberg.  The
earliest plate imaging VES 263 that was found in them is from 1896, the
latest was exposed in 1995, thus encompassing a whole century of the
object's history.

With the assistance of Edward Los of the Harvard College Observatory we
accessed, prior to public release, the results of DASCH scans (Grindlay et
al.  2012) of Harvard plates imaging the region containing VES 263.  After
culling the large number of untrustable plates, we found a total of 206
reliable DASCH measurements of VES 263 on Harvard plates which are plotted
in Figure~3.  Two of us (S.Y.S.  and R.J.-S.) have directly accessed
the plate archives of the Sternberg Astronomical Institute (SAI) in Moscow
and of the Astronomical Observatory in Sonneberg (Germany) and looked for
plates imaging VES 263.  A total of 155 good blue-sensitive plates were
found and measured (82 from the SAI 40cm astrograph, and 73 from various
Sonneberg astrographs).  The results are given in Table~4 and included in
Figure~3.  The agreement between Harvard, Moscow and Sonneberg plates is
excellent in view of the limitations inherent in photographic plates and
object variability.  The three sets are complementary, with the Harvard
plates covering the earlier years better, while the Moscow and Sonneberg
plates extend to more recent epochs, and in particular filling-in the
so-called Menzel's gap (1955-1965) that affected the Harvard sky patrol. 
The formal error of the photographic plate measurements ranges from
0.05 to 0.2 mag (median value 0.1 mag), while the dispersion of the
individual datapoints along the mean lightcurve in Figure~3 
is $\sigma$=0.11 mag.

The photometric behaviour in Figure~3 suggests VES 263 to have lingered
around the quiescence $B$-band level ($B$=14.93 according to APASS) until
about 1945 when the star begun a slow rise from $B$$\sim$15.05 reaching
$B$$\sim$14.15 by $\sim$1953.  It then remained around $B$$\sim$14.10 until
about 1969 when it begun a descent reaching $B$$\sim$14.85 around 1972. 
This was followed by a resumption in brightness increase peaking at
$B$$\sim$14.60 by 1984 and a new descent to $B$$\sim$14.85 by the time of
the last photographic images exposed in 1995, with a trend suggesting a
return to quiescence level by $\sim$1998.

The average $B$$\sim$14.10 level during the 1953-1969 {\it plateau} is far
brighter than the $B$$\sim$14.65 value reached at the peak of the current
2018-19 outburst.  For amplitude and duration, the 1953-1969 plateau is
the largest event in the recorded photometric history of VES 263.  On the
other hand the rise to it has been slow and gradual, and the present 2018-19
eruption could be the start of a new activity cycle similar to that leading
to the plateau of half a century ago.

For completeness we report that in the Sternberg plate archive in Moscow we
found a further 32 plates going just deep enough to have barely recorded VES
263.  These blue-sensitive plates were exposed between 1896 and 1947 with
two smaller astrographs located in Moscow, with lenses 10 and 16 cm in
diameter.  On these plates VES 263 is seen close to the faint limit of the
plates, which precludes an accurate measurement of its brightness. 
Nonetheless, these plates are useful to show that VES 263 was always close
to quiescence brightness during this whole period, and thus excluding any
major brightening like the one recorded during the 1953-1969 plateau.

\subsection{Spectroscopic behaviour}

A sample of the low resolution optical spectra that were  acquired for VES
263 during the 2018-29 eruption are presented in Figure~4 corresponding to
minimum, maximum and mid brightness.  The very red slope of the continuum is
obvious as is the presence of a strong H$\alpha$ line in emission as also
the signatures of insterstellar absorption (atomic lines and DIBs).  Some
HeI lines are seen in weak emission, 5876 and 7065 \AA\ being the strongest.

    \begin{figure*}
    \includegraphics[width=160mm]{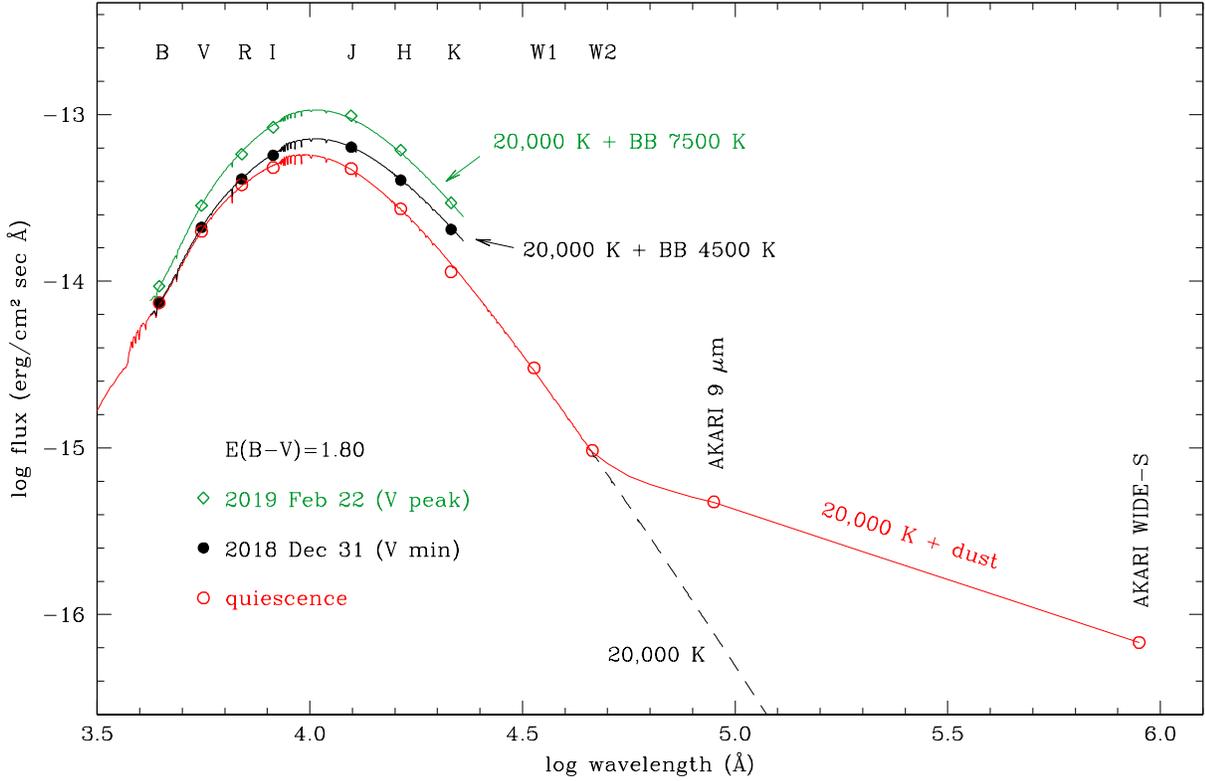}
    \caption{The spectral energy distribution of VES 263 at three distinct
    states (quiescence, and minimum and peak brightness during the current
    eruption). The SED in quiescence is fitted with a synthetic spectrum
    for the B1II star and black-body emission over 400$-$25~K for the dust.
    Blackbodies are added to the B1II star for the other epochs (all reddened 
    by the same $E_{B-V}$=1.80).}
    \end{figure*}

A medium resolution optical spectrum of VES 263 in eruption is presented in
Figure~5, focusing on the classical 3850$-$5000~\AA\ interval used for
spectral classification.  CP12 rated their B1II classification (performed
over the same 3850$-$5000~\AA\ interval) as having an uncertainty negligible
in the spectral type (0.5 subtypes) and larger (2 classes) in luminosity
class.  Consequently, in Figure~5 we compare our spectrum of VES 263 with
that of template stars for spectral types B1III, B1II and B1I, selected from
the Yamashita \& Nariai (1977) list of recommended standards.  They have
been observed the same night with the same telescope and instrumental
configuration as used for VES 263.  For an easier comparison, their slopes
have been forced to replicate that of the much more reddened VES 263, and
have off-sets added for plot clarity.  The most significant insterstellar
atomic lines and DIBs are marked.

An immediate difference with the spectrum presented by CP12 (their Figure~2)
concerns the H$\beta$ line.  The CP12 spectrum was obtained in quiescence
and presents a H$\beta$ in full absorption, while the eruption spectrum in
Figure~5 is characterized by a filled-in H$\beta$ with a double-peaked
emission profile which is plotted on a velocity scale in the insert.  In
addition, the fine details of the absorption lines appear different from
CP12.  The redder the continuum becomes, the less pronounced are the
absorption lines, as if they are veiled by continuum emission from a cooler
source.  This is quite obvious when comparing the intensity of HeI 4026 with
that of HeI 4922 in VES 263 and in the template stars.  The appearance of
HeI 4922 is significantly weakened in VES 263, while HeI 4026 stands much
closer to the intensity displayed in the template stars.  The putative
cooler source, whose emission veils weaker lines from the B1II central star,
seems also to add broader wings to the Balmer lines.  Such broad wings are
absent in the spectra of the template stars and also in the quiescence
spectrum shown by CP12 (even if the limited resolution of their plot could
lead to a wrong impression here).

Just five days past the medium-resolution optical spectrum presented in
Figure~5, we have obtained the Infrared spectrum shown in Figure~6 and
covering the $Y$, $J$, and $H$ wavelength intervals.  There is no trace of
the B1II photospheric absorption lines, the spectrum being dominated by the
continuum emission from the cooler source responsible for the photometric
eruption.  The Paschen and Brackett series of hydrogen appear in emission as
also a fairly strong HeI 1.083 $\mu$m feature.  Weaker features at 1.0686
and 1.1287 $\mu$m are attributed to CI and OI.  Since the OI 1.1287 $\mu$m
line is significantly strengthened by  Lyman continuum fluorescence (Mathew
et al.  2018) and it is seen here in emission whereas the continuum
fluoresced OI 1.3164 $\mu$m is not, it implies the presence of a copious
source of LyC photons whose source must obviously be the hot B1II 
central star.  Comparing the slope of the optical spectra in Figure~4 rising
toward the red, with that of the Infrared spectrum in Figure~6 going the
opposite way, it's evident how the peak of VES 263 observed spectral energy
distribution is placed around 1.0 $\mu$m.

It is also worth noting that the Brackett lines in the $H$-band are
optically thick because  Br10 at 1.7362 $\mu$m is weaker than the higher
Brackett lines like Br11, Br12 etc.  whereas it would be expected to be
stronger had Case B conditions been followed for emission coming from  an
optically thin gas.  That the HI lines are optically thick is not a
surprise, since as shown in coming sections, the object is identified as a
HAeBe star with a disk and the disk in such stars, from which the HI lines
originate,  are expected to have high densities of 10$^{11}$ to 10$^{13}$
cm$^{-3}$ (Mathew et al.  2018, and references within).  At such high
densities the Brackett hydrogen lines are expected to be optically thick
(Hummer \& Storey 1987; Storey \& Hummer 1995)

    \begin{figure}
    \includegraphics[width=85mm]{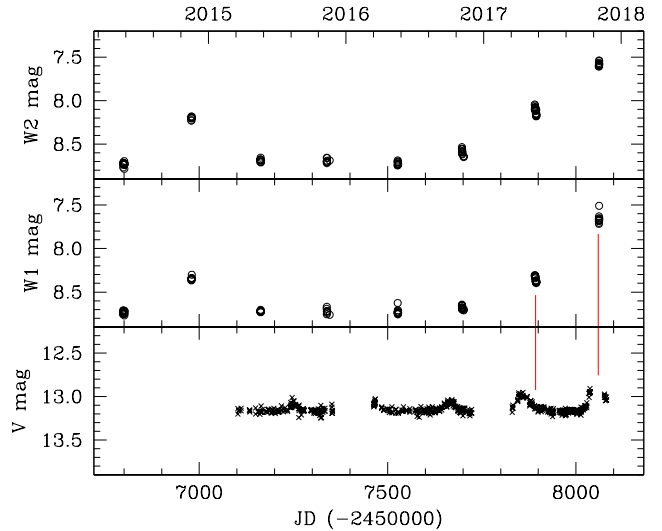}
    \caption{Comparison between NeoWISE data and $V$-band photometry of VES
    263. The lines are meant to guide the eye.}
    \end{figure}

\section{The spectral energy distribution}

The spectral energy distribution of VES 263 is presented in Figure~7.  Three
epochs are considered: those of minimum and maximum in brightness during the
current 2018-19 eruption, and the phase of preceeding quiescence.

The spectral energy distribution (SED) in quiescence is built from 2009-2014
multi-epoch APASS optical photometry (reporting average values as $B$=14.93,
$V$=13.15, $R$=12.11, and $I$=11.22), 2MASS $J$$H$$K_s$ data for 1998 (Cutri
et al.  2003), AKARI S9W,WIDE-S measurements for 2006-2007 (Ishihara,
et al.  2010; Ali-Lagoa et al.  2017), and AllWISE\footnote{We ignore
AllWISE W3,W4 data because of conflicting results reported for PSF-fitting
and aperture photometry measurements, and their large scatter depending on
the radius of aperture.} W1,W2 data for 2010 (Cutri et al.  2013). Such
data are obviously non-simultaneous, but since the object was in a quiescent
state during this time, it should result in any significant difference.

The calibrations of the MK spectral types in terms of photospheric
temperature, summarized by Tokunaga (2000) and Drilling \& Landolt (2000),
average to 20,000~K for B1 giant/supergiants.  To fit the energy
distribution for VES 263 in quiescence in Figure~7, we have selected from
the grid computed by Castelli and Kurucz (2003) the synthetic spectrum
appropriate for a B1II star (parameters: $T_{\rm eff}$=20,000~K,
$[$M/H$]$=0.0 and $\log g$=3.0), and adopted the Fitzpatrick (1999)
extinction law for the standard $R_V$=3.1 case.  The excellent fit in
Figure~7 is obtained for $E_{B-V}$=1.80.  A similar good fit and
reddening would be obtained adopting a blackbody distribution for
$T$=20,000~K.  Both fitting with a synthetic spectrum or with a blackbody
are not over-sensitive to temperature: considering that the optical-IR is
located on the Rayleigh-Jeans tail of the energy distribution for stars as
hot as the B1II in VES 263, the actual temperature makes a small
difference (fits with 19,000 or 21,000 K blackbodies or synthetic spectra
would perform equally good in Figure~7), and the shape of the observed SED
is actually controlled by the reddening, with a change by
$\Delta$$E_{B-V}$=$\pm$0.05 being readily appreciable.  

The SED in quiescence shows a large excess over the 20,000~K
photospheric emission for $\lambda$$\geq$6~$\mu$m.  The IR-excess in
Figure~7 is fitted by adding black-body emission from dust distributed in
temperature from 400 to 25~K.  The presence of warm circumstellar dust is a
distinctive feature of HAeBe stars (eg.  Stahler \& Palla 2005).  The
luminosity radiated by such dust is $L_{\rm IR}$$\sim$12~L$_\odot$.  It was
obtained by integrating the (unreddened) flux in excess of the 20,000~K
photospheric synthetic spectrum in Figure~7, and scaling it to the distance
of VES 263 (sect.  6).

The SED of VES 263 at minimum and peak brightness during the 2018-19
eruption are plotted in Figure~7 using our simultaneous $B$$V$$R$$I$ and
$J$$H$$K$ observations for 2018 Dec 31 and 2019 Feb 22, respectively.  A
good fit to the SED at minimum is obtained by adding a 4500~K blackbody to
the synthetic spectrum for the B1II star.  The SED at maximum is instead
fitted by adding a 7500~K blackbody to the emission from the B1II central
star (both 4500~K and 7500~K blackbodies being reddened by the same
$E_{B-V}$=1.80 affecting the 20,000~K synthetic spectrum).  Such 4500~K
and 7500~K blackbodies have, respectively, $\lambda$-integrated luminosities
of 120 and 860 L$_\odot$.  This fitting exercise shows that the brightness
and colors during the eruption of VES 263 are governed by the presence and
variability of a source both cooler and separate/additional to the central
B1II star, that we identify with the circumstellar disk (see sect.8 below)
that is a typical feature of pre-Main Sequence objects (Stahler \& Palla
2005).  The surface temperature of a disk obviously depends on the distance
from the central star, and therefore the 4500~K and 7500~K values found
above are to be interpreted as averages weighted over the respective
surfaces emitting at $B$$V$$R$$I$ and $J$$H$$K$ wavelengths.

It is worth noticing that the minor brightenings visible in the pre-eruption
lightcurve of Figure~2, are also linked to the variable presence of a source
much cooler than the stellar photosphere (i.e. the disk).  This can be easily
inferred from Figure~8, where the NeoWISE data (Mainzer et al.  2011, 2014)
available for VES 263 are plotted in phase with the $V$-band lightcurve.  It
is evident how (1) during quiescence the NeoWISE $W1$ and $W2$ data (Nugent
et al.  2015) stay flat and close to the AllWISE $W1$=8.58 and $W2$=8.50 mag
values characterizing the quiescent SED of Figure~7, and (2) any minimal
brightening visible in the $V$-band reverberates into a much larger increase
at NeoWISE $W1$ and $W2$ wavelengths.  The AllWISE catalog combines
observations from the 2009-2010 cryogenic and post-cryogenic survey phases
of the NASA Wide-field Infrared Survey Explorer (WISE), and NeoWISE refers
to the data the satellite is collecting since it has been brought out of
hibernation and resumed observation in 2014.

\section{Astrometric membership to the Cyg OB2 association}

Analysis of {\it Gaia} astrometry supports VES 263 being a member of
the Cyg OB2 association.  To this aim, we started by acquiring a list of hot
OB stars in this association compiled by Wright, Drew \& Mohr-Smith (2015)
and in an extended region around the densest part of the association by CP12
and Berlanas et al.  (2018).  As their membership designation is
historically mainly attributed by over-density of spectroscopically
confirmed OB stars in that region (Ivanov 1996), we first queried {\it Gaia}
DR2 data (Gaia Collaboration et al.  2018) in a cone with the radius of
1$^\circ$ centered at $\alpha$~=~$308.163^\circ$ $\delta$~=~$41.299^\circ$
and matched the {\it Gaia} observations with previously determined members. 
Already confirmed by Berlanas et al.  (2019), the members are concentrated
around parallax $\sim$0.58~mas with a possible extension to closer
distances.  VES 263, marked with a solid vertical line in Figure~9, is
located close to the peak of that distribution.

        \begin{figure}
	\centering
	\includegraphics[width=85mm]{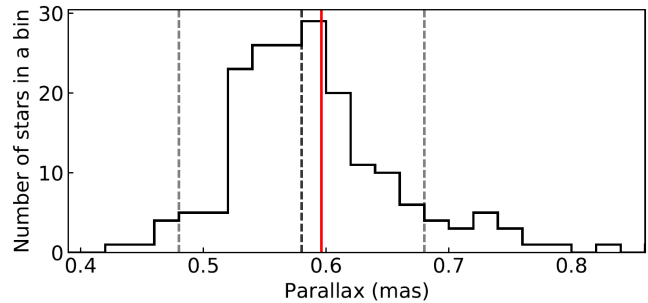}
	\caption{Distribution of {\it Gaia} DR2 parallax measurements of
	confirmed members of the Cyg OB2 association.  The parallax of VES
	263 is marked with the red vertical solid line.  Black vertical
	dotted lines indicate our parallax selection, centred at 0.58~mas,
	that was used to study the proper motion of the association.}
	\end{figure}
	\begin{figure}
	\centering
	\includegraphics[width=85mm]{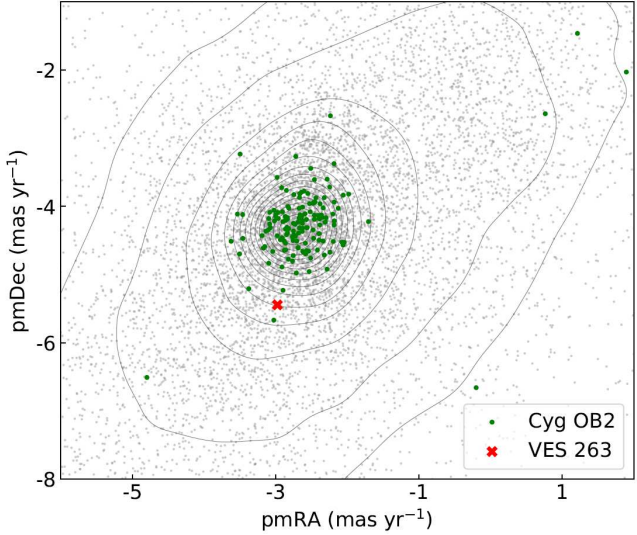}
	\caption{Distribution of selected stars in proper motion space, with
	indicated association members and VES 263.  The density profile,
	shown as equidistant contours, was created by summing 2D Gaussian
	probability density function (PDF) of all stars.  PDFs were defined
	by stellar proper motion and its reported uncertainty.}
	\end{figure}
	\begin{figure}
	\centering
	\includegraphics[width=85mm]{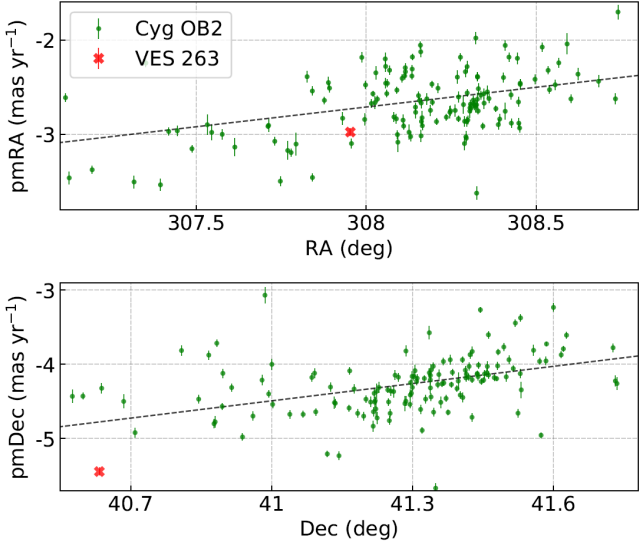}
	\caption{Distribution of proper motion and sky location of the Cyg
	OB2 members and VES 263 in the right ascension ({\it top}) and
        declination ({\it bottom}) planes.}
	\end{figure}

Having more than $800,000$ stars, that could smear out over-densities of few
hundreds of stars in the proper motion space, we limited our selection to
stars with a parallax measurement between $0.48$ and $0.68$~mas (also
indicated in Figure~9).  This selection resulted in 11561 stars of which 153
are members of the association, together plotted in Figure~10.  The same
figure shows over-density of stars with the same proper motion as known Cyg
OB2 members.  If we move our selection to higher/lower parallax values the
over-density disappears.  VES 263 falls on the outskirt of this peak, where
the density of stars in proper motion space starts to increase. 
Disregarding the clear velocity outliers, that might not be part of the
association, VES 263 has a velocity consistent with other members.  As the
OB associations are proven not to be bound (Mel'nik \& Dambis 2017; Wright
2018) and have larger physical sizes than open clusters with considerable
small-scale kinematics substructure (Wright 2018), a greater velocity
scatter is expected for them.  As VES 263 is not located in the main part of
the association, but in its extended region, its velocity vector might
indicate radial expansion.  Plots in Figure~11 show dependency of members'
proper motion on their position in the sky.  Both plots show some trends,
where VES 263 is consistent with the scatter along the fitted line and
therefore membership to Cyg OB2.  

From fitting to isochrones Wright et al.  (2015) concluded that the majority
of star formation in Cyg OB2 occurred more or less continuously between 1
and 7 Myr ago, which should then bracket the age of VES~263.  At an
estimated total mass of $\sim$16,500~M$_\odot$, Cyg OB2 is one of the most
massive groups of young stars known in our Galaxy.

\section{Distance and Interstellar Reddening}

The Gaia DR2 parallax for VES 263 is 0.5962 mas, with an error of just
$\pm$0.0253 (4\%) that allows a safe direct inversion (Gaia Collaboration et
al.  2018) to derive the distance as $\sim$1.68$\pm$0.07 kpc.  At a Galactic
latitude of only $+0.60$ deg, the sightline accumulates a lot of
interstellar reddening upon reaching VES 263, as already clear from the SED
fitting of the previous section requiring $E_{B-V}$=1.80$\pm$0.05.

    \begin{figure*}
    \includegraphics[width=170mm]{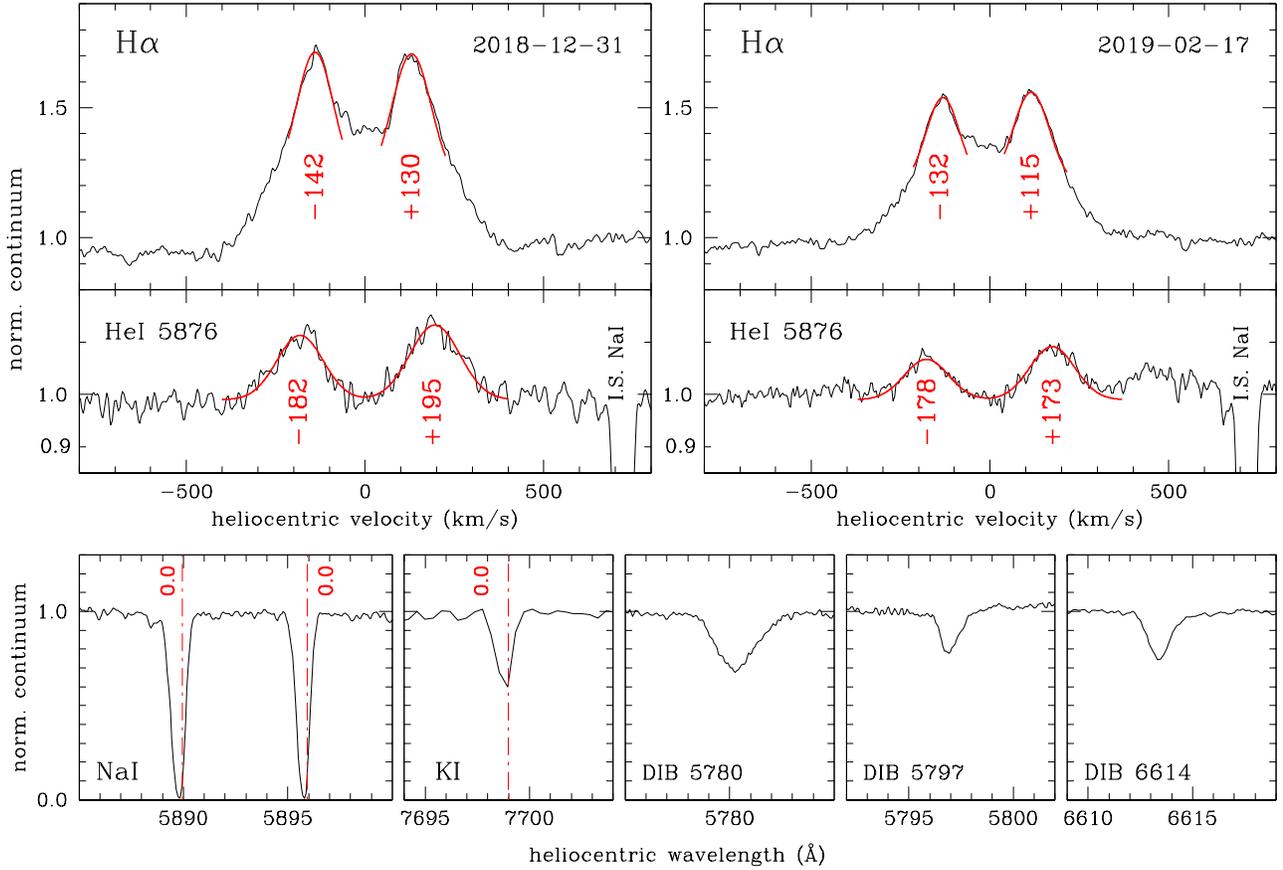}
    \caption{{\it Top}: Gaussian fitting to the double-peaked line profiles
    of H$\alpha$ and HeI 5876 for minimum (left) and maximum (right)
    brightness during the 2018-19 eruption. {\it Bottom}: interstellar lines
    and DIBs as recorded on the Asiago Echelle spectra. The dash-dotted
    lines mark null heliocentric radial velocity.}
    \end{figure*}

There are several 3D maps of interstellar extinction that may be employed
to derive estimates of the reddening affecting VES 263.  {\sc Stilism} by
Lallement et al.  (2014) and Capitanio et al.  (2017) provides
$E_{B-V}$=1.98$\pm$0.25, and the 3D Pan-STARRS1 map by Green et al.  (2018)
returns $E_{B-V}$=2.06$\pm$0.21.  An appreciable lower error is associated
to the IPHAS 3D map by Sale et al.  (2014) giving an extinction $A_V$=6.19,
with lower and upper limits at $A_V$=5.85 and $A_V$=6.55 combining IPHAS and
Gaia uncertainties.  For the $R_V$=3.1 extinction law, using the quadratic
expression of Fiorucci and Munari (2003) for early B-type stars suffering
from high reddening, this translates to $E_{B-V}$=1.92 with formal lower and
upper limits of  $E_{B-V}$=1.82 and $E_{B-V}$=2.03.

    \begin{table}
    \caption{Summary of results from reddening indicators. E.W. is the
    equivalent width of the interstellar spectral feature and $E_{B-V}$ 
    the corresponding reddening as derived from the published relations
    discussed in sect. 6.}
    \includegraphics[height=82mm]{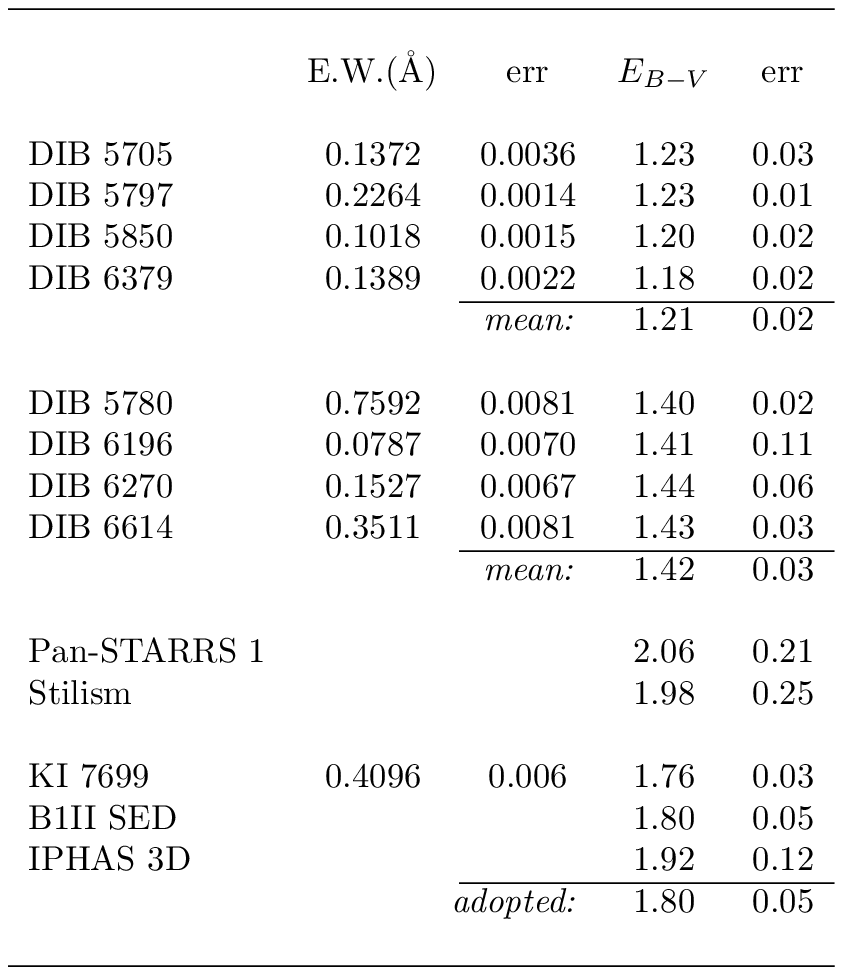}
    \end{table}

At such  high reddening, the interstellar NaI lines are too strong (core
saturated) in our Echelle spectra of VES 263 to allow using them to derive
the reddening, as illustrated in Figure~12.  Interstellar KI lines are
however still far from saturation.  The equivalent width 0.410$\pm$0.006
\AA\ that we have measured for KI 7699 \AA\ on our Echelle and medium
resolution spectra (cf.  Figure~12) translates into $E_{B-V}$=1.76$\pm$0.03
following the calibration by Munari and Zwitter (1997).

The diffuse interstellar bands (DIBs) do present a general trend to get
stronger with increasing reddening (e.g.  Jenniskens \& Desert 1994), but
the correlation is far from being a tight or unique one.  Our Echelle
spectra of VES 263 have recorded a rich sample of DIBs, a few being shown in
Figure~12.  Two sightline categories, named $\zeta$ and $\sigma$, have been
described by Krelowski et al.  (1992) and Krelowski \& Sneden (1994) which
correspond to {\em UV-shielded} and {\em non-shielded} sightlines, probing
cloud cores and external regions, respectively.  Vos et al.  (2011) show
that there are fundamental differences between the two groups in
correlations between DIBs, reddening, and gas.  The two sightline categories
are named after those toward $\zeta$ Oph and $\sigma$ Sco.  Kos and Zwitter
(2013) found that DIBs at 5705, 5780, 6196, 6202, and 6270 \AA\ do not show
much difference between $\zeta$ and $\sigma$ sightlines, while DIBs at 4964,
5797, 5850, 6090, 6379, and 6660 \AA\ have distinctly different relations
for $\zeta$ and $\sigma$ sightlines between their equivalent width and
$E_{\rm B-V}$ reddening.  The ratio of the equivalent width (EW) of DIBs
5780 and 5797 allows to distinguish between the two types of sightlines: an
EW$_{5797}$/EW$_{5780}$$>$0.3 corresponds to $\zeta$ category, and
conversely an EW$_{5797}$/EW$_{5780}$$<$0.3 to $\sigma$ category.  The DIBs
at 5780 and 5797 \AA\ in our Echelle spectra of VES 263 are shown in
Figure~12.  Their equivalent widths are 0.755 and 0.330 \AA\, respectively,
corresponding to $\zeta$-type sightline categories.  In Table~5 we report
our measurement of the equivalent width of the DIBs best visible in our
Echelle spectra of VES 263, and the corresponding reddening $E_{\rm B-V}$
derived by applying the $\zeta$-type calibration given by Kos and Zwitter
(2013).  Such reddenings tightly cluster around two distinct values, $E_{\rm
B-V}$=1.21$\pm$0.02 and $E_{\rm B-V}$=1.42$\pm$0.03.  Such values are
mutually incompatible and also far too small compared to all other
indicators summarized in Table~5.  Other E.W./$E_{\rm B-V}$ calibrations for
DIBs (e.g.  Munari 2014) result in values similar to those of Table~5, and
therefore we conclude that DIBs along the sightline to VES 263 do not
conform to published reddening relations calibrated over large portions of
the sky.

The three most accurate reddening determinations among those summarized in
Table~5 come from KI 7699, SED fitting and IPHAS 3D map, for an average
value of $E_{\rm B-V}$=1.80$\pm$0.05 that we will adopt in the rest of this
paper.

\section{Absolute magnitude and the HR diagram}

From the Gaia DR2 parallax, $E_{\rm B-V}$=1.80 reddening and APASS
$V$=13.15, the absolute magnitude of VES~263 in quiescence is $M_V$=$-$3.75
mag.  This is equivalent to a bolometric magnitude $M_{\rm bol}$=$-$5.64 mag
adopting from Flower (1996) a bolometric correction B.C.=$-$1.89 for
giants/super-giants with $T_{\rm eff}$=20,000~K, and $M^{Sun}_{\rm
bol}$=+4.74 (Bessell, Castelli \& Plez 1998; Torres 2010).  The
corresponding luminosity is $L$=14,000 $L_\odot$, which is close to
$L$=12,000 $L_\odot$ as obtained by direct flux integration of the
(unreddened) synthetic $T_{\rm eff}$=20,000~K spectrum in Figure~7.

The position on the HR diagram of HAeBe stars as catalogued by Vioque
et al.  (2018) is shown in Figure~13.  The position and error bars for
VES~263 correspond to $L$=13,000$\pm$3,000 $L_\odot$ and $T_{\rm
eff}$=20,000$\pm$2,000~K. The corresponding blackbody radius would be 
9.5$\pm$1.2 $R_\odot$.  In the same figure we have plotted the pre-Main
Sequence tracks for three values of the mass from Bressan et al.  (2012),
leading to an estimate of $\approx$12~M$_\odot$ for VES 263.

\section{The accretion disk}

\subsection{Evidence from emission line profiles}

The emission profile for H$\alpha$ and HeI 5876 in VES 263 at the time of
maximum and minimum brightness during the 2018-19 eruption are plotted in
Figure~12.  Such double-peaked profiles are typical of circumstellar disks
(Horne \& Marsh 1986).  The origin in a disk is reinforced noting that the
peaks of HeI lines have a larger velocity separation than H$\alpha$.  The
excitation potential for HeI 5876 is $\sim$23 eV, twice larger than the
$\sim$12 eV for H$\alpha$, and population of the upper level for HeI 5876
requires hotter and denser regions compared to H$\alpha$, which means an
inner disk radius and therefore a faster Keplerian rotation.  The ratio of
velocity separation of the double-peaked H$\alpha$:HeI~5876 profiles goes
like 1.0:1.4 for VES 263, which quite favorably compares with the average
value 1.0:1.5 we have measured for a sample of quiescent cataclysmic
variables (CVs) we have observed over the years with the same Asiago Echelle
spectrograph used to observe VES~263.  In CVs, the emission lines are well
known to form in the accretion disk around the central white dwarf (e.g. 
Warner 1995).

From line fitting in Figure~12, the projected rotational velocity of the
disk at the location of H$\alpha$ formation is $v \sin i$$\sim$130
km~s$^{-1}$.  Assuming a pure Keplerian rotation for the disk and adopting
from the previous section 12~M$_\odot$ and 9.5$\pm$1.2 $R_\odot$
for the central star, the H$\alpha$ forms at $\approx$14 stellar radii for
an inclination of the disk $i$=90$^\circ$ and $\approx$7 stellar radii for
$i$=45$^\circ$.

    \begin{figure}
    \includegraphics[width=85mm]{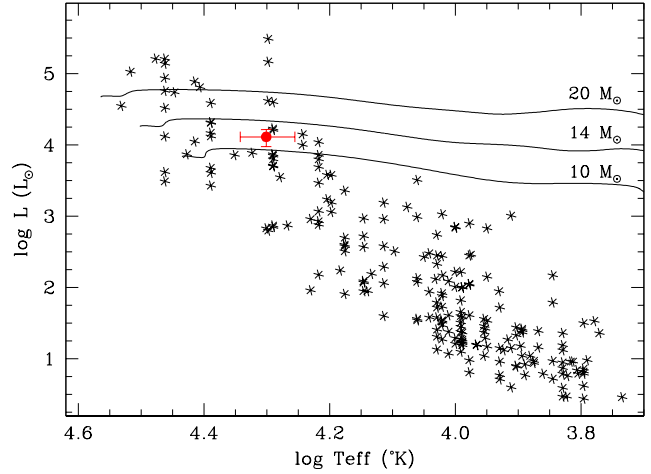}
    \caption{The position of VES 263 on the HR diagram. The asterisks mark
    data for the HAeBe stars in the compilation by Vioque et al.  (2018).
    The lines are pre-Main Sequence evolutionary tracks from Bressan et al.
    (2012).}
    \end{figure}

In addition to the SED distributions in Figure~7, the disk as the agent
reponsible for the variability observed in 2018-19 is also supported by the
evolution with time of the integrated flux of H$\alpha$, which is listed in
Table~2 and plotted in Figure~14.  The H$\alpha$ behaviour in Figure~14
matches exactly the lightcurve of VES 263.  For the latter we selected the
$I$-band brightness because its effective wavelength is the closest to the
peak of the energy distribution of VES 263 (cf.  Figure~7).  The $\Delta
I$=0.50 mag amplitude of the lightcurve translates into a change of the
photometric flux by 58\%, which is the same as the 54\% change in the
integrated flux of H$\alpha$ as listed in Table~2.

    \begin{figure}
    \includegraphics[width=85mm]{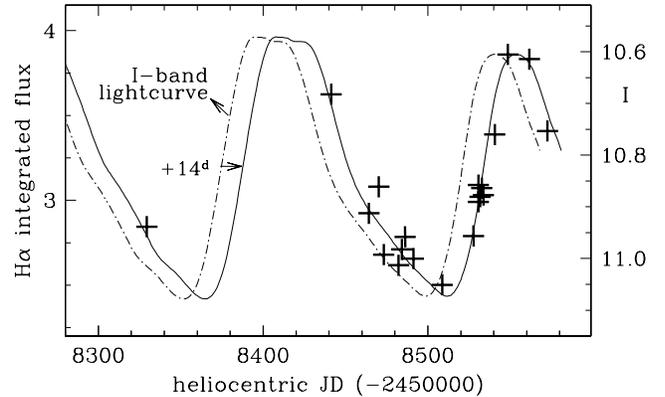}
    \caption{Evolution with time of the integrated flux of H$\alpha$
    emission line of VES 263, from data in Table~2.  The dot-dashed line is
    the $I$-band lightcurve from Figure~1, the same shifted by 14 days is
    plotted as the continuous line.}
    \end{figure}

A noteworthy feature in Figure~14 is the $\sim$14~days shift between
H$\alpha$ and the $I$-band.  The brightness of the disk is proportional to
the mass-flow through it (Ichikawa \& Osaki, 1992).  Similarly, the
integrated flux of H$\alpha$ traces the mass-flow through the inner regions
of the disk where the temperature is high enough to excite the emission from
the line.  The $\sim$14 days delay appears therefore to be the average time
required for mass to migrate from the outer regions of the disk where the
bulk of the emission in the $I$-band originates to the inner regions where
the temperature is hot enough to excite emission from H$\alpha$.  

A final remark on the disk is offered by Figure~12. Close to minimum
brightness (Dec 2018 spectrum) the velocity separation of the double-peaked
line profiles is larger than at maximum brightness (Feb 2019 spectrum).  A
higher brightness corresponds to a larger mass flow through the disk, and
therefore a given temperature is reached at an outer radius (given the
classical $T_{\rm eff} \propto M_{\rm acc}^{1/4}$ dependence).  The
temperature required to excite emission from H$\alpha$ is then reached at a
greater distance from the central star, resulting in a lower Keplerian
velocity and consequently a reduced velocity separation of the double-peaked
line profile (similarly for HeI 5876).  Precisely what is seen in Figure~12.

\subsection{Looking for orbital motion}

The high resolution Echelle spectroscopic observations of VES 263 listed in
Table~2 cover a time interval of about 200 days.  Photospheric absorption
lines are veiled by featureless continuum emission by the disk, leaving only
the H$\alpha$ emission line to derive the radial velocity.  A sample of the
recorded H$\alpha$ profiles is presented in Figure~15.

Many massive stars are found in binary systems, as it is the case for the
majority of O- and B-type stars in open clusters and OB associations (e.g. 
Boeche et al.  2004), with mass ratios closer to $q\sim$1 than less massive
binaries.  Their orbital periods can be as short as 1-2 days with
consequently orbital velocities of hundreds of km~s$^{-1}$.  The line
photocenter in Figure~15 is stable around the average value of
$RV_\odot$=$-$4.1$\pm$0.9 km~s$^{-1}$.  The dispersion of measurement
($\sigma$=2.5 km~s$^{-1}$) is similar to the dispersion of intra-night
individual spectra, suggesting minimal or null change in radial velocity
during the monitored 200 days.  This leads to basically three different
explanations: (1) the B1II star is single, or any companion is far less
massive ($q\ll$1), (2) the orbital period is much longer than the sampled
200 days, or (3) the orbital inclination is very low (face-on orientation). 
The latter explanation is the less probable, because it is reasonable to
assume that in a binary the circumstellar disk lies close to the orbital
plane, and the disk in VES 263 has a considerable inclination given the
large velocity separation of the peaks in the emission line profiles
(Figure~12).  Alternatives 1 and 2 could either be true, as their
combination.  To further investigate them, we plan to continue gathering
high resolution spectra of VES 263 in the future.

In addition to keeping their photocenter stable, the line profiles in
Figure~15 do not change their shape other than reducing the velocity
separation in response to changes in the mass flow through the disk.  This
indicates that over the 200 days of monitoring the inner regions of the disk
where H$\alpha$ forms remained circular and symmetric, excluding precession
of an asymmetric shape for the disk.

    \begin{figure}
    \includegraphics[width=85mm]{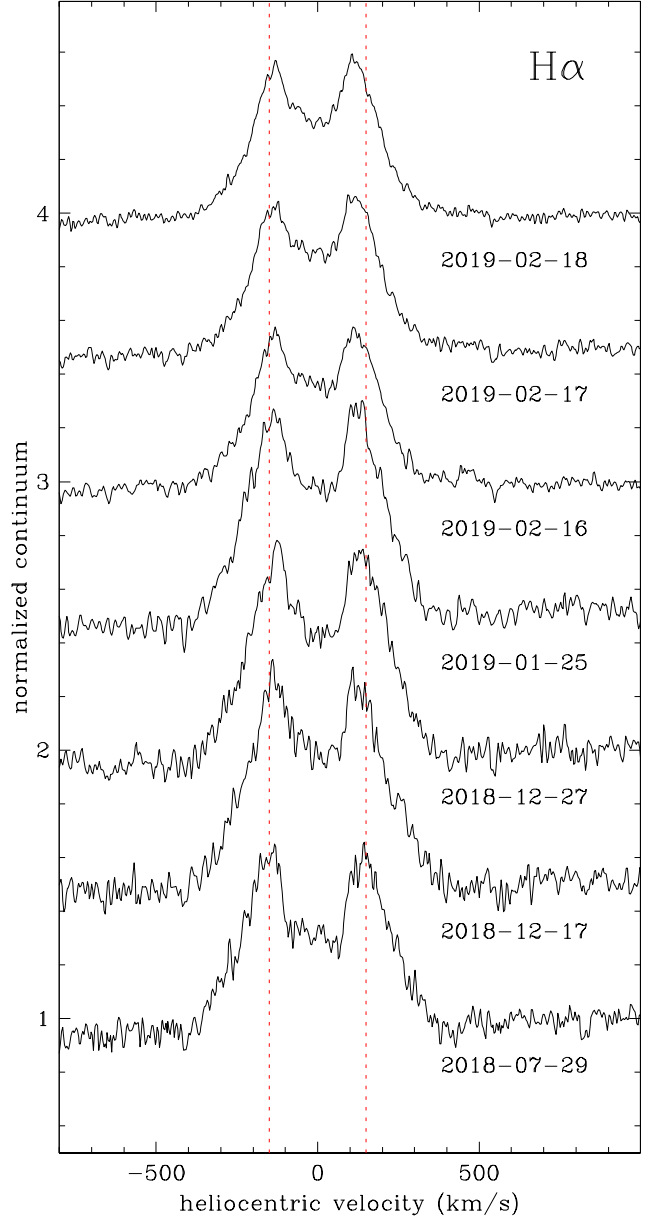}
    \caption{Comparison of the H$\alpha$ emission profile of VES 263
    as observed at different epochs with the Asiago 1.82m + Echelle
    spectrograph (adjacent continuum normalized to 1.0). The lines
    at $-$150 and $+$150 km~s$^{-1}$ are plotted to guide the eye.}
    \end{figure}

\subsection{Mass accretion rate}

Mass accretion ($M_{\rm acc}$) and accretion luminosity ($L_{\rm acc}$) are
related by
\begin{equation}
L_{\rm acc} = G \frac{M_\ast M_{\rm acc}}{R_\ast}
\end{equation}
where $M_\ast$ and $R_\ast$ are the mass and radius of the accreting object,
respectively. 
This can be used to obtain an order-of-magnitude estimate of the mass 
flow rate through the disk of VES~263 and onto its central star. 

The SED of VES 263 at maximum and minimum brightness during the current
2018-19 eruption is characterized by excess radiation at optical and near-IR
wavelengths that, when fitted with blackbodies and their
emission integrated over the whole wavelength range, provides luminosities
of 860 and 120 L$_\odot$, respectively (cf. Figure~7 and sect.4).  The
corresponding mass accretion rates would be 1.12$\times$10$^{-5}$ and
1.43$\times$10$^{-6}$ M$_\odot$~yr$^{-1}$.  They are however lower limits to
the actual value of $M_{\rm acc}$, because the fit with blackbodies of
$T_{\rm eff}$=7,500 and 4,500 K miss the emission from the inner and
hotter regions of the disk able to sustain the formation of Balmer and HeI
emission lines.  To access them it would be essential to gather satellite
observations in the ultraviolet, difficult to obtain in view of the
very large excintion affecting VES~263 (of the order of 15 mag at 2000
\AA).

To test the hypothesis that the accretion luminosities derived in the
previous paragraph are lower limits, we may estimate $L_{\rm acc}$ from the
reddening-corrected observed flux ($F_{\rm line}$) of emission lines,
following a common practice for pre-Main Sequence objects.  The isotropically
radiated luminosity $L_{\rm line}$ in the given line is related to distance
$d$ through
\begin{equation}
L_{\rm line} = 4 \pi d^2 F_{\rm line}
\end{equation}
The transformation of $L_{\rm line}$ into $L_{\rm acc}$ is usually obtained
by power-law relations of the type
\begin{equation}
\log (\frac{L_{\rm acc}}{L_\odot}) = A_{\rm line} 
+ B_{\rm line}\times \log(\frac{L_{\rm line}}{L_\odot})
\end{equation}
with calibrations of $A_{\rm line}$,$B_{\rm line}$ coefficients existing for
many emission lines.  Integrating the flux of emission lines visible in the
near-IR spectrum of Figure~6 and adopting their $A_{\rm line}$,$B_{\rm
line}$ coefficients from Fairlamb et al.  (2017) leads to accretion
luminosities ranging from 200 to 1300~L$_\odot$, with a median value of
900~L$_\odot$.  This confirms as a lower limit the 120 L$_\odot$ accretion
luminosity derived above from integration of SED for the same date (2018 Dec
31), the time of minimum during the current 2018-19 eruption.  It would have
been interesting to repeat the exercise with a similar near-IR spectrum
taken during a maximum in the current 2018-19 eruption, but unfortunately we
have none.  We do have however in Table~2 an extended series of flux
measurements for the H$\alpha$ emission line.  Adopting the corresponding
$A_{\rm line}$,$B_{\rm line}$ coefficients from Mendigut{\'{\i}}a et al. 
(2011), the derived accretion luminosities range from 280 to 450~L$_\odot$. 
The amount of reddening correction to H$\alpha$ is however vastly
larger than for the near-IR emission lines, and uncertainties on $E_{B-V}$ could
easily account for part of the 2$\times$ difference in their respective $L_{\rm
acc}$.

The accretion luminosity $L_{\rm acc}$ has also been found to correlate
with the luminosity of the central star $L_{\star}$.  The analysis of
existing data by Mendigut{\'{\i}}a et al.  (2015), shows that for HAeBe the
accretion luminosity ranges from $L_{\rm acc}$$\sim$$L_{\star}$ to
0.01\,$L_{\star}$.  VES~263 follows this rule with -- from data above --
$L_{\rm acc}$$\sim$0.07$\times$$L_{\star}$ at minimum and $L_{\rm
acc}$$\sim$0.5$\times$$L_{\star}$ at maximum of the current 2018-19
eruption.

Finally, it appears worth noticing that the recent calibration by Arun et
al.  (2019) of $M_{\rm acc}$ as funtion of the mass of the central star in
HAeBe objects predicts $M_{\rm acc}$$\approx$3.3$\times$10$^{-5}$
M$_\odot$~yr$^{-1}$ for a 12~M$_\odot$ star as in VES 263.  This well
compares with the lower limit $M_{\rm acc}$$\geq$1.1$\times$10$^{-5}$
M$_\odot$~yr$^{-1}$ estimated in this section for the accretion rate at the
peak of the present VES 263 eruption.

    \begin{figure}
    \includegraphics[width=85mm]{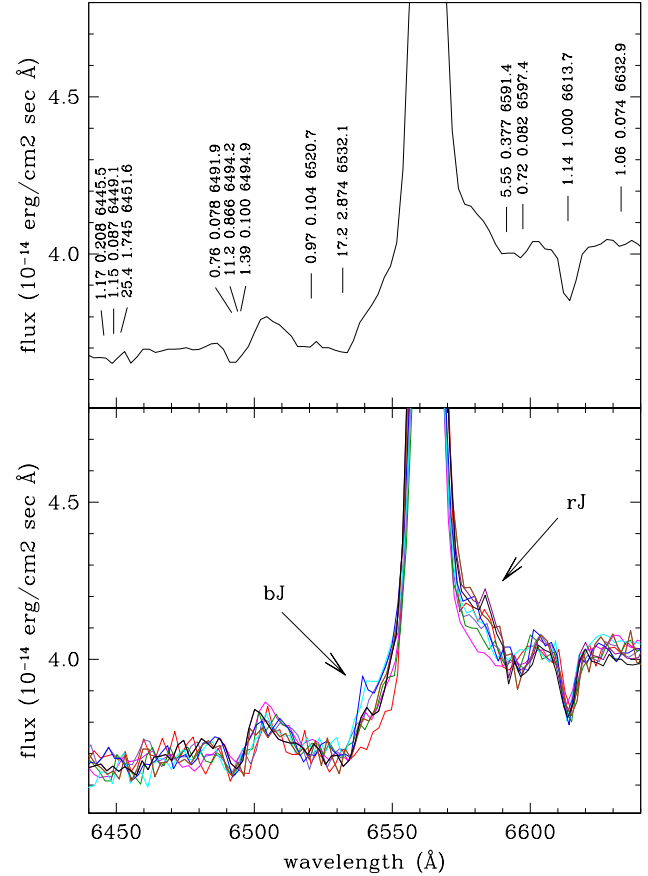}
    \caption{{\it Top panel:} Average around H$\alpha$ of the low-res
    spectra of VES 263.  The strongest DIBs in the compilation by Jenniskens
    \& Desert (1994) are marked (see sect. 8.4  for associated numbers).  {\it
    Bottom panel:} the individual spectra making up the above average are
    plotted individually (after matching their median value).  The arrows
    point to regions at $\approx$ $-$700 and $+$700 km~s$^{-1}$ where the 
    dispersion is larger (see sect. 7.4).}
    \end{figure}

\subsection{Wind and jets}

The low resolution spectra listed in Table~2 show a low-level variability in
the immediate vicinity of H$\alpha$ emission line.  They are compared in
Figure~16.  To avoid confusion with the many DIBs present in the region, the
principal ones are identified in the top-panel of the figure.  Three numbers
are given for each DIB, namely the FWHM (in \AA), the equivalent width (in
\AA, scaled to that of DIB 6614), and the wavelength of the photocenter (in
\AA), averaging among the values reported by Jenniskens \& Desert (1994) in
their surveys of DIBs toward several hot stars. Figure~16 shows that indeed
the dispersion of individual spectra is larger in the immediate
vicinity of H$\alpha$.

A variable {\it hump} is present in emission to the red of H$\alpha$, at a
bulk velocity of $\approx$ +700 km~s$^{-1}$.  Symmetrically placed at a bulk
velocity of $\approx$ $-$700 km~s$^{-1}$ there is a variable absorption, or as an 
alternative a variable second emission hump.  To distinguish between these
two alternatives for the blue feature is not easy in view of the limited
spectral resolution and S/N of the spectra in Figure~16.  To complicate the
matter is the deep depression of the stellar continuum caused by the nearby
very strong DIB 6532.  Devoted observations, at higher S/N and greater
spectral dispersion, are required to properly address the issue, which we
plan to perform in the future.  These features are relatively minor, carrying
a flux $\sim$1/10 of that of the main H$\alpha$ line.  They are barely at the
threshold of detection in Echelle observations, being lost in the strong
curvature imposed by the instrumental blaze function at the center of the
order, where H$\alpha$ is situated.

We propose the features at $+$700 km~s$^{-1}$ and $-$700 km~s$^{-1}$ to be
the signature of possible bipolar jet ejection from the central regions of
the accretion disk surrounding the central star.  That at $-$700 km~s$^{-1}$
could appear in absorption instead of emission if seen projected onto the
central B1II star or the brightest inner regions of the disk.  The escape
velocity from the central star is $\sim$700 km~s$^{-1}$ (assuming $R$=9.5
R$_\odot$ and $M$=12 M$_\odot$).  The similarity of the jet velocity and the
escape velocity is a general property of all objects known to posses
collimated mass-outflows (Livio 1997).  Bipolar jets have been observed in
many HAeBes (Stahler \& Palla 2005), sometimes extending to such great
distances from the central star to be easily resolved spatially by
ground-based observations (e.g.  G{\"u}nther, Schneider, \& Li, 2013;
Melnikov et al.  2008; Grady et al., 2004; Corcoran \& Ray, 1998).

\section{Conclusions}

The observational  evidence discussed in previous sections proves the
variable presence of a massive accretion disk in VES 263.  The SED in
Figure~7 shows how the contribution of such a disk was negligible when the
object was in quiescence, and become dominant at redder wavelengths during
the current 2018-19 eruption.  Furthermore, the strict parallel behavior
between disk brightness and integrated H$\alpha$ flux shown in Figure~14
proves how the ups-and-downs in the lightcurve are correlated to the
variable mass-flow through the disk.

What causes such a variability in the mass being fed to the disk is unknown,
and its investigation is beyond the scope of the present paper.  It is
however intriguing to note from Figures~1 and 2 how the photometric activity
and mean brightness have been increasing on the average during the last few
years, signalling a gradual resumption of mass-feeding and flow through the
disk.  This behavior seems parallel to the slow and contrasting rise
($\sim$8 yrs) from quiescence to peak brightness at the time of the large
eruption of the 1950's, illustrated in Figure~3.  It is tempting to argue
that the 1950's eruption was caused by a similar resumption of the accretion
disk as we are witnessing now.  If the current event should turn into a
replica of the eruption of the 1950's, then we are at present only just
half-way to its peak brightness.

VES 263 is clearly worth a continuing detailed, multi-wavelength monitoring
over the coming years, as well as a dedicated effort to locate further
information about its colorful past history.

\section{Acknowledgements}

We are grateful to the anonymous referee for valuable comments that
helped to improve the paper. We acknowledge Jonathan Grindlay and Edward
Los of Harvard College Observatory for their permission to use DASCH data on
VES 263 prior to inclusion in public DR6 scheduled for 2019.  R.J.S.  thanks
the team of Sonneberg Observatory and 4pi Systeme company for hospitality
and Peter Kroll and Eberhard Splittgerber for help and permission to use the
historical plate archive.  The research work at the Physical Research
Laboratory is funded by the Department of Space, Government of India. 
S.Y.S.  is supported by Grant VEGA No 2/0008/17 and Development Program from
M.V.  Lomonosov Moscow State University ('Physics of stars, relativistic
objects and galaxies').  K.{\v C}.  acknowledges financial support of the
Slovenian Research Agency (research core funding No.  P1-0188 and project
N1-0040).  R.J.-{\v S}.  is supported by University of Rijeka under the
project number uniri-prirod-18-3.  The observations presented in this paper
have been collected in part with ANS Collaboration telescopes, the Galileo
1.22m telescope of the University of Padova, and the INAF Copernico 1.82m
and Schmidt 67/92cm telescopes.

\end{document}